\documentclass[aps,prb,twocolumn,superscriptaddress]{revtex4-2}
\usepackage{CJKutf8}
\usepackage{graphicx,xcolor}

\begin{document}
\begin{CJK*}{UTF8}{gbsn}
\title{Spatial mapping of magnetization fluctuations via time-lapse Kerr microscopy}

\author{X.-Y. Ai(艾显越)}
\email[]{xianyue.2.ai@uni-konstanz.de}

\affiliation{Department of Physics, University of Konstanz, 78457 Konstanz, Germany}
\author{I. V. Soldatov}
\affiliation{Leibnitz Institute for Solid State and Materials Research Dresden, Helmholtzstr.20, D01069 Dresden, Germany}

\author{L. Oleschko}
\affiliation{Department of Physics, University of Konstanz, 78457 Konstanz, Germany}

\author{Seema}
\affiliation{Department of Physics, University of Konstanz, 78457 Konstanz, Germany}

\author{M. Mueller}
\affiliation{Department of Physics, University of Konstanz, 78457 Konstanz, Germany}

\author{S. Karpitschka}
\affiliation{Department of Physics, University of Konstanz, 78457 Konstanz, Germany}

\author{R. Schäfer}
\affiliation{Leibnitz Institute for Solid State and Materials Research Dresden, Helmholtzstr.20, D01069 Dresden, Germany}
\affiliation{Institute for Materials Science, TU Dresden, 01062 Dresden, Germany}

\author{S. T. B. Goennenwein}
\affiliation{Department of Physics, University of Konstanz, 78457 Konstanz, Germany}

\date{\today}

\begin{abstract}
Combining wide-field magneto-optical Kerr microscopy with a time-lapse analysis scheme allows investigating magnetization fluctuations with high spatial as well as temporal resolution. 
We here use this technique to study magnetization fluctuations in a thin ferromagnetic film prepared into a quasi-equilibrium magnetic state via a dedicated field sweep protocol. 
Our experiments reveal spatially localized noise hotspots distributed across the sample surface within the magnetic domains in the quasi-equilibrium state. 
The spatial density of the noise hot spots is very similar at different low magnetic field strengths, as expected from the fluctuation-dissipation theorem. The measurement scheme thus opens the way for the spatially resolved investigation of quasi-equilibrium noise processes in magnetic materials.
\end{abstract}
\maketitle
\end{CJK*}

\section{Introduction}

Magnetic domains have attracted interest from both a fundamental perspective, and regarding potential applications in spintronic memory and logic devices \cite{parkin1,memory2,memory3}. 
As magnetic domain formation takes place on mesoscopic scales, the statistical properties of magnetic domains have been addressed in  experimental \cite{MOnoi3,statex1,statex2} and theoretical \cite{stattheo1,stattheo2,stattheo3} investigations.
Owing to the fluctuation dissipation theorem (FDT) \cite{FDTbook}, the study of magnetization fluctuations  has emerged as a powerful tool in the investigation of magnetic materials.  
Indeed, magnetization fluctuations detected via magneto-resistive \cite{MRes1,MRes2,MRes3,MRes4} or magneto-optical \cite{MOnoi2,MOnoi3,MOnoi1} measurements have provided valuable insights into, e.g., the physics of spin ice systems or spin-reorientation transitions.

Despite these advancements, the observation of stochastic magnetization fluctuations in equilibrium -- quantitatively covered by the FDT \cite{FDTbook,Kogan} -- has proven elusive for conventional ferromagnets \cite{ocio_86}. 
According to the FDT, large fluctuations are expected when the magnetic susceptibility $\chi=\partial M/ \partial H$ is large \cite{FDTbook}. At the same time, the system should be in (quasi-)equilibrium, such that the magnetization expectation value is time-independent. These conditions are simultaneously met only in particular systems such as spin ice \cite{ocio_86} or thin films engineered to feature anhysteretic magnetic properties \cite{MOnoi2}. In contrast, in conventional ferromagnetic thin films with an open (hysteretic) $M(H)$ loop -- i.e., with two different magnetization states $M$ for one given external magnetic field strength $H$ -- the occurrence of a large magnetic susceptibility typically is accompanied by magnetic relaxation effects, in which the magnetization expectation value irreversibly evolves with time \cite{Wiegman1978, rudibuch_domaineffect}. 
More often than not, large magnetization fluctuation signals thus arise in experiments performed in a non-equilibrium magnetic configuration, such that the validity of the FDT is jeopardized.

Vice versa, it is well established that slow magnetic relaxation or magnetic creep is facilitated by a thermally activated process, and that the  fraction of nucleated reverse domains then expands in small abrupt steps \cite{durin2023, rudibuch_domaineffect}. Those events are sometimes referred to as Barkhausen steps because of their similarity with the step-wise magnetization changes induced by a changing external magnetic field, reported by Barkhausen more than a century ago \cite{barkhausen}.
Even though such a step-wise magnetization change historically was associated with a type of magnetic noise, it is  "not amenable to a thermodynamic analysis since it is essentially an off equilibrium effect" as pointed out by Ocio et al.~\cite{ocio_86, rudibuch_domaineffect}.  To mitigate this issue,  long wait times before the start of fluctuation measurements have been used in some experiments, in order to let the system relax into a state close to equilibrium before taking data \cite{MOnoi4}. 

In this paper, we use a different approach to study magnetization fluctuations in quasi-equilibrium in ferromagnetic Co/Pt/AlOx thin film samples with perpendicular magnetic anisotropy (PMA): We intentionally prepare the sample into a quasi-equilibrium state via dedicated field sweeps \cite{JILES}, and then use 
wide-field magneto-optical Kerr microscopy (MOKE microscopy) for a spatially resolved study of the ensuing magnetic fluctuations. 
More specifically, for a given constant magnetic field magnitude, we record series of MOKE microscopy images as a function of time. This yields a data set similar in structure   
to the one employed by Durin et al.~in Ref.~\cite{durin2023}. 
Analyzing these data  allows for a spatially-resolved study of the magnetic noise response in quasi-equilibrium. 
Surprisingly, we find that the spatial distribution of the magnetization noise is highly non-uniform. The density of local noise 'hot spots' is large when the slope $\chi$ of the anhysteretic magnetization loop is large, as expected from the FDT. Moreover, the position of the noise hot spots appears to be correlated with the emerging domain structure. 
We interpret our results as evidence that domain wall dynamics are of key importance for the stochastic magnetization fluctuations in ferromagnetic thin films.
Moreover, the strongly localized characteristic of the noise observed indicates that the complex interaction between magnetic domain structure and material pinning sites plays a crucial role, as also reported in confined magetnic structures \cite{eltschka}.

\section{Experimental details}
\begin{figure}
\includegraphics[scale=1]{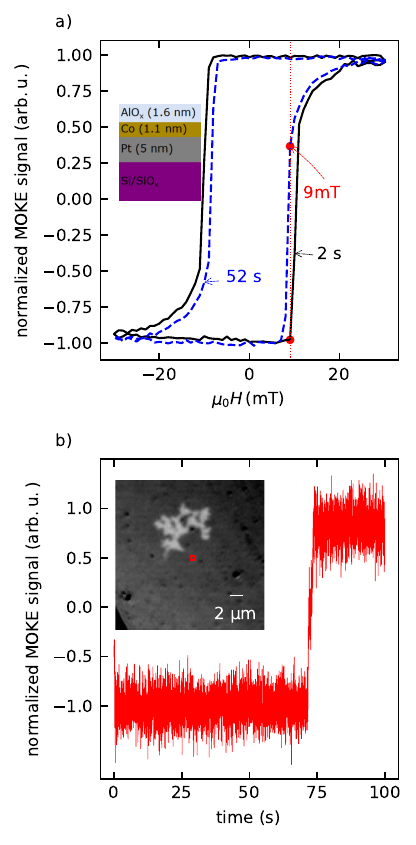}
\caption{\label{hysteresis}(a) Magnetic hysteresis loop of the AlOx/Co/Pt trilayer sample, obtained by calculating the average  intensity $I_{\mathrm{avg}}$ of the MOKE microscopy image (and normalizing it to $\pm 1$) for each magnetic field applied. The black line shows data recorded with a new magnetic field value every two seconds, while the data represented by the dashed blue line were recorded with $50\,\mathrm{s}$ additional delay between subsequent field values. Inset: Schematic of the layer stack. (b) Time-lapse Kerr signal of the pixel marked by the red square in Kerr microscopy image shown in the inset. The time-lapse data were taken for a fixed magnetic field value of $+9~\mathrm{mT}$ (see the red line in panel (a)) over a time period of $100\,\mathrm{s}$, after initially saturating the sample at large negative fields. }
\end{figure}

We use a AlO$_x$(1.5nm)/Co(1.1nm)/Pt(5nm) trilayer sputter-deposited on a Si/SiO$_x$ substrate for the experiments. A schematic representation of the layer sequence is depicted in the inset of Fig.~\ref{hysteresis}a). The sample exhibits strong perpendicular magnetic anisotropy (PMA), as evidenced by the rather square shape of the magnetic hysteresis loop recorded in MOKE measurements with the external magnetic field applied along the surface normal (Fig.~\ref{hysteresis}a)). All measurements discussed in this publication were performed at room temperature.

For the magneto-optical noise measurements, we use a commercial Kerr microscope. The system is operated in p-MOKE mode, such that it is only sensitive to component of the magnetization $M_\perp$ oriented along the surface normal (the out-of-plane magnetization component) \cite{rudibuch_technik, selective}. 
The intensity of the light reflected from the sample and passing through the polarization analyser is described by Malus' law as $I_{\mathrm{Kerr}}=k_0 I_0 \cos^2(\alpha)$ \cite{advanced_kerr}. Here, $k_0$ is the
transmittance of the analyzer, $I_0$ is the intensity of the light before entering the analyzer and $\alpha$ is the angle of polarization rotation from magneto-optical effects. 
For small polarization rotation angles and large enough opening of the analyser, 
this dependency can be approximated by the linear relation $I_{\mathrm{Kerr}}=a \cdot M_\perp$ \cite{advanced_kerr,soldatov_quantitative,RAVE_quantitative,rudibuch_technik}. Here, $a$ stands for the conversion factor between the light intensity $I$ (i.e., gray level in the Kerr microscope camera image) and the out-of-plane magnetization  $M_\perp=m_\perp / V$. 
$V$ represents the volume of the sample probed by the Kerr measurement, and $m_\perp$ is the out-of-plane magnetic moment in this volume.  
To achieve high spatial resolution, we used a 50x objective lens with a numerical aperture $\mathrm{NA}=0.8$. 
This results in an optical resolution of approximately $0.5\,\mathrm{\mu m}$ in the  Rayleigh limit for the $640\,\mathrm{nm}$ wavelength of the light emitting diodes used in the microscope \cite{ivan_resolution}.

For Kerr image acquisition, a Hamamatsu ORCA C11440-36U monochrome sCMOS camera proved ideal due to  its high pixel well depth (PWD) of over 30000 electrons (compared to a few thousand for other state of the art cameras \cite{orca_quest}), which minimizes shot noise scaling with light intensity as $S_{\mathrm{shot}} \propto \sqrt{I}$ while still delivering reasonable frame rate.
In simple terms, the impact of shot noise on the quality of the data thus is reduced when the image is obtained by collecting more light \cite{rudibuch_technik}. 
The camera yields images with $960 \times 540$ pixels using $2\times 2$ binning at the device level, providing a spatial resolution of approximately $0.23\,\mathrm{\mu m}$ per pixel. 
This is about half of the optical resolution limit, implying that the sample surface areas represented by each pixel overlap to some extent.  
We therefore chose a 3x3 averaging of neighboring pixels for the time resolved analyses as a good compromise between spatial resolution and shot noise mitigation.

Time resolution is achieved via so-called time lapse analysis, i.e., by taking series of consecutive MOKE images, one every 20 milliseconds, for fixed externally applied magnetic field magnitude.  
The data transfer and storage capacity of the computer system controlling the Kerr microscope allows to record such image series for at least 100 seconds at 50 fps.
Each raw image is corrected for spatial drift via image registration \cite{pyramid}. 

Vice-versa, averaging over all pixels in a given Kerr image (or appropriate regions of interest if the sample is micropatterned)  yields an average MOKE signal value $I_{\mathrm{avg}}$. Plotting the latter as a function of the applied magnetic field strength allows to straightforwardly record  magnetic hysteresis loops. 
Figure \ref{hysteresis}(a) shows a typical example. In this figure, the $I_{\mathrm{avg}}$ values have been normalized to $+1$ ($-1$) at large positive (negative) external field, respectively.

\section{Fluctuation dissipation theorem and magnetic relaxation}
The fluctuation-dissipation theorem (FDT) establishes a fundamental connection between fluctuations within an extensive quantity and its susceptibility towards the conjugate intensive quantity \cite{FDTbook,Kogan}.
For ferromagnetic systems, the FDT can be expressed as \cite{MRes1,Kogan}
\begin{equation}
    S_m(f) = \frac{2k_B T}{\pi \mu_0 f} \mathrm{Im}(\chi_m(f)),
    \label{fdto}
\end{equation}
with the noise power density $S_m (f)$, the Boltzmann constant $k_B$, the temperature $T$, and the vacuum permeability $\mu_0$. Note that the susceptibility $\chi_m=\chi \cdot V$  here is expressed in terms of the magnetic moment $m=M\cdot V$ in the volume $V$ of interest.

Furthermore, the real and imaginary parts of $\chi_m$ are interrelated through the Kramers-Kronig relation \cite{FDTbook,Kogan}
\begin{equation}
\mathrm{Re}(\chi_m(f=0)) = \frac{2}{\pi} \int_{0}^{\infty} \frac{\mathrm{Im}(\chi_m(f))}{f}\, df 
\label{kramers}
\end{equation}
Using $I_{\mathrm{Kerr}}=a \cdot M_\perp=\frac{a}{V}\cdot m_\perp$ as discussed above, the total magnetic noise power can be expressed in terms of the Kerr signal intensity as \cite{MOnoi2, Kogan}

\begin{equation}
\int_{0}^{\infty} S_{\text{Kerr}}(f) \, df = \frac{k_B T}{\mu_0 V} ~a~\frac{\partial I_{\mathrm{kerr}}(f=0)}{\partial H}
\label{eq-SKerr}
\end{equation}

From a theory perspective, Eq.~(\ref{eq-SKerr}) elegantly connects noise power and susceptibility. 
Experimentally, however, relaxation effects often jeopardize the determination of the relevant DC (zero frequency) susceptibility. 
The two magnetic hysteresis loops depicted in Fig.~\ref{hysteresis}(a) illustrate this issue. 
The loops were recorded one with $2\,\mathrm{s}$, one with $50\,\mathrm{s}$ wait time, respectively, at each magnetic field value before data acquisition. 
Longer wait times -- or equivalently a slower field sweep rate -- obviously result in a narrower width of the hysteresis.  
This implies that the susceptibility is wait-time dependent, although the data points were taken at nominally the same,  constant field. 
The noise power according to Eq.~(\ref{eq-SKerr}) then also is ambiguous. 
This is very prominently so for $\mu_0 H=9\,\mathrm{mT}$ in Fig.~\ref{hysteresis}(a): depending on the wait time, the sample is either still saturated (corresponding to a small susceptibility and thus presumably a small noise power), or is in the reversal process with a large susceptibility. 

Time-lapse Kerr microscopy experiments reveal that the magnetic relaxation evident from Fig.~\ref{hysteresis}(a) is due to domain creeping taking place on the time scale of seconds.  
Such slow magnetic relaxation effects are often observed in PMA thin film samples \cite{durin2023,grassi, creflow, rudibuch_domaineffect}.
Notably, for small sample volumes like the one represented by an individual pixel in the Kerr microscopy image, abrupt switches of the local magnetic moment will occur at constant external field. Figure \ref{hysteresis}(b) exemplarily shows corresponding data. 
The sample thus clearly is not in an equilibrium state, and the FDT is not applicable \cite{othernoise1,othernoise2,MRes1}. 
\section{Experimental preparation of quasi-equilibrium magnetic states}
\begin{figure}
\includegraphics[scale=1]{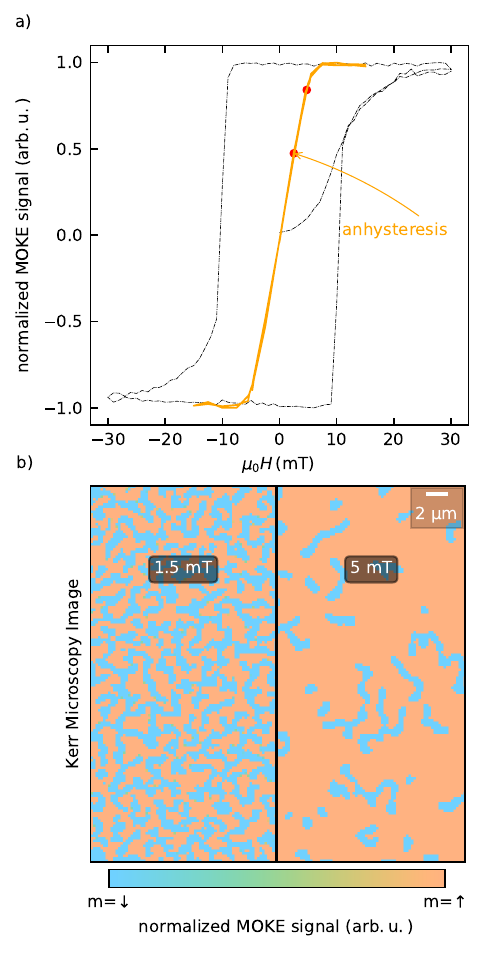}
\caption{\label{anhysteresis}(a) Using a combination of DC and AC magnetic fields, the sample is brought into a quasi-equilibrium magnetic state (orange line). This state is anhysteretic, in contrast to the conventional magnetic hysteresis loop (dashed black line). (b) Kerr Microscopy images of the quasi-equilibrium magnetic states of our PMA sample, recorded at room temperature at two different DC magnetic field strengths.}
\end{figure}

In order to investigate the quasi-equilibrium magnetic noise properties as a function of field strength, we prepared the sample into a quasi-equilibrium magnetic state \cite{old}. To this end, a given DC magnetic field $H_0$ applied to the sample is superimposed by an additional AC field $H_{\mathrm{AC}}$. We used an AC field frequency of 1 Hz and an amplitude of 30 mT. For DC field strengths $|\mu_0 H_0 | \lesssim 15\,\mathrm{mT}$, this ensures that the sample still is magnetically saturated at both the positive and negative extrema of the AC field oscillation. 
The AC field amplitude $H_{\mathrm{AC}}$ is then  linearly decreased to zero over a period of 50 seconds. At the end of the process, the sample thus again is residing in the original DC magnetic field $H_0$ only. 
Repeating this preparation protocol for different values $H_0$ of the DC field yields the magnetization loop shown by the orange line in Fig.\,\ref{anhysteresis} a). 

Importantly, this loop is anhysteretic - the DC magnetic field sweep direction or more generally the previously applied magnetic fields do not impact the magnetic state at a given $H_0$. 
The anhysteretic magnetization loop thus yields a field-dependent quasi-equilibrium state of the sample, free from domain creeping and other time-dependent relaxation effects \cite{anhysteresis}. 

However, we would like to stress that determining or experimentally reaching the magnetic domain state with the lowest energy is challenging. The energy differences between various possible domain patterns typically are too small to be distinguishable \cite{rudibuch_domaineffect}. Therefore we refer to the domain state achieved after applying the specific field sweep protocol described above as "quasi-equilibrium", since it very likely is not the lowest energy one. 

\section{Noise analysis in magnetic quasi-equilibrium}
\begin{figure}
\includegraphics[scale=1]{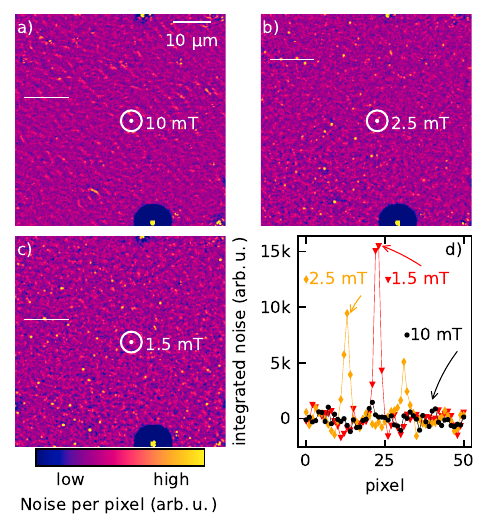}
\caption{\label{hotspots}Spatial maps of the integrated noise power (calculated  from the time lapse data) reveal the appearance of local noise hot spots in the magnetization reversal regime. (a) In magnetic saturation at $10\, \mathrm{mT}$, the spatially resolved noise power only shows structure corresponding to the roughness of the sample surface. (b),(c) In the magnetization reversal region, at $2.5\, \mathrm{mT}$  and $1.5\, \mathrm{mT}$, respectively, localized noise hot spots (evident as bright yellow speckles) appear in the same portion of the sample. Panel (d) depicts position-dependent integrated noise along the white lines shown in (a)-(c), illustrating that the noise hot spots have significant but non-uniform noise power and spatial distribution. The lines in (a) and (c) are at the same location, the line in (b) is chosen to cut two noise hot spots.}
\end{figure}

Conventional Kerr microscopy experiments show magnetic domain structures in the quasi-equilibrium magnetic states, as depicted in Fig.\,\ref{anhysteresis} b). The domains are randomly distributed across the sample, and the relative fraction of up and down domains depends on the external field magnitude. 

The time-lapse Kerr microscopy technique described above enables the study of the time evolution of local magnetization in the quasi-equilibrium state. Applying the preparation protocol detailed in the previous section for a given DC field $H_0$ and then taking  time-lapse Kerr microscopy data yields the spatial and temporal evolution of the magnetic state for this magnetic field strength. 
Repeating this type of experiment for a series of different $H_0$ allows to follow the domain evolution and the changes in the local magnetic noise power density across the entire quasi-equilibrium magnetization reversal process.
Figure \ref{hotspots} shows corresponding data. As evident from the figure, we find that the noise power density (calculated using the Wiener-Khinchin theorem as the Fourier transform of the $I(t)$ autocorrelation function \cite{Kogan}) is strongly enhanced for certain groups of pixels. This effect is directly obvious in the data after applying unsharp-masking \cite{Burger2022}, which eliminates the global spatial inhomogeneity in the noise power caused by inhomogeneous illumination.
 
As also shown Fig.\,\ref{hotspots}, the distribution of noise hot spots (the density of bright yellow speckles) qualitatively changes with $H_0$. 
In particular, a substantial density of hot spots is present only when the sample is \textit{not} saturated. The noise hot spots thus must be connected to the magnetic domain structure, and cannot be simply attributed to shot noise resulting from intensity differences between pixels or other experimental artifacts such as mechanical stability. 
One exception are defects on the surface that strongly reflect light, which results in a much higher noise level. The larger bright spot at the bottom of Fig.\,\ref{hotspots}(a)-(c), slightly to the right of the middle, is an example for such structural artefacts in the noise maps. 

Within the magnetization reversal region, however, the density of noise hot spots appears to be mostly independent of field strength, while their positions clearly vary with $H_0$. For a unit surface of around $600\,\mathrm{\mu m}^2$, the ratio of binarized noisy pixels to 'conventional' ones is $0.82 \%$ for $2.5\, \mathrm{mT}$ and $0.99 \%$ for $1.5\,\mathrm{mT}$, respectively. In saturation at $10\, \mathrm{mT}$, this 
ratio comes down to $0.30 \%$, which we attribute to the finite density of highly reflecting surface defects. 
Some of those points present at saturation are noisy enough to be also visible at $1.5\,\mathrm{mT}$ and $2.5\, \mathrm{mT}$. 
Qualitatively, this is consistent with the FDT: The anhysteretic $M(H)$ evolution (see Fig.\,\ref{anhysteresis}) has approximately constant slope $\partial M/\partial H$ and thus constant (large) susceptibility in the reversal field range. The magnetic noise thus also should be large.
In contrast, at $\mu_0 H_0 =10\,\mathrm{mT}$, the sample is magnetically saturated. The susceptibility then is much smaller, with a corresponding noticeable decrease in noise hot spots. 

\begin{figure}
\includegraphics[scale=1]{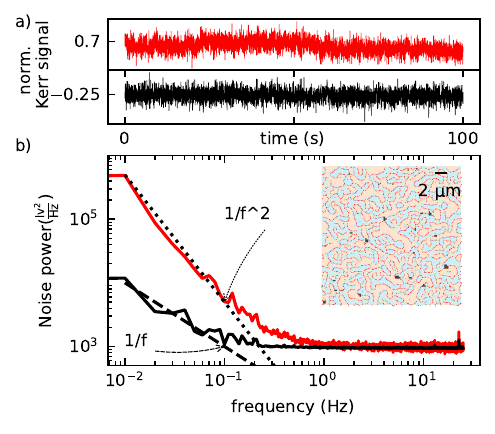}
\caption{\label{equilibrium} (a) Normalized time traces of two individual Kerr microscopy image pixels at $2.5\, \mathrm{mT}$. The upper pixel (red line) is a 'noisy' one (see the main text for details), the lower one is 'conventional'. (b) Noise power density spectra calculated from the data set including time traces in (a) at $2.5\, \mathrm{mT}$. The inset shows a heat-map of the noise distribution over a small region of the sample at $2.5\, \mathrm{mT}$, which is used to group the pixels for the averaged noise power density spectra.}
\end{figure}

To corroborate the quasi-equilibrium nature of the magnetic state prepared via the field sweep protocol, we show the time evolution of the Kerr signal intensity $I(t)$ of two typical pixels in Fig.\,\ref{equilibrium}(a) - one from a noise hot spot, one not.  
For both pixels, the Kerr intensity fluctuates with time as expected considering the shot-noise limited optical detection process. 
Importantly, there is no discernible change towards saturation in the average intensity, in stark contrast to the situation in Fig.\,\ref{hysteresis}(b). 
The sample is therefore indeed in a quasi-equilibrium magnetic state. 
Nevertheless, as depicted in Fig.\,\ref{equilibrium}(b), we  find that the averaged pixels in the noise hot spots (red line) have a low-frequency noise scaling as $S(f)\propto 1/f^\beta$ with $\beta \approx 2$, while the conventional pixels outside the hot spots (black line) feature $\beta \approx 1$.

Furthermore, the presence of noise hot spot sites seemingly correlates with the curvature of domain walls.
The inset of Fig.\,\ref{equilibrium}(b) displays the binarized noisy pixels (now in black) in a small region on the sample at $2.5\, \mathrm{mT}$, superimposed on top of the magnetic domain structure. The noise hot spots seem to be located at or very close to magnetic domain walls. 
While a detailed analysis is beyond the scope of this paper, this observation underscores the intricate interplay between thermal effects, domain wall dynamics, and magnetic energy landscape in determining the overall behavior of magnetic materials still to be explored. Remarkably, our observations are in line with the findings reported in \cite{eltschka}, which suggest that domain walls exhibit thermally activated motions around pinning sites. 

\section{Conclusions}
Combining a time-lapse analysis scheme with state-of-the-art wide-field Kerr microscopy, we have studied the magnetization noise in quasi-equilibrium in a thin magnetic film with perpendicular magnetic anisotropy. Our approach enables both high spatial as well as temporal resolution. Since magnetic relaxation effects jeopardize the applicability of the fluctuation-dissipation theorem, we used a dedicated magnetic field sweep protocol to prepare the magnetic system into a quasi-equilibrium state. 
Within the magnetization reversal region, we find strongly localized magnetic noise hot spots. The local enhancement of magnetic fluctuations appears to be connected with the magnetic domain structure, possibly in analogy to the domain state instability seen in magnetically anhysteretic systems such as ferromagnets near the spin reorientation transition at zero fields \cite{MOnoi3}.
More detailed work in the future will be necessary to fully elucidate this point. 
Nevertheless, our results show that magnetization fluctuations in quasi-equilibrium prevail also in conventional ferromagnetic thin films with an open hysteresis loop. Our approach can be straightforwardly transferred to the investigation of fluctuations in other magnetic systems and magnetic domains close to equilibrium. 
\begin{acknowledgments}
We thank G. Sala for the sample provided, M. A. Weiss, F. Herbst and N. Nabben for fruitful
discussions and input. This work was financially supported
by the Deutsche Forschungsgemeinschaft (DFG, German
Research Foundation) via the Collaborative Research
Center SFB 1432 (Project no. 425217212).
\end{acknowledgments}

\end{document}